\documentclass[11pt]{article}

\usepackage{amsmath}
\usepackage{amssymb}
\usepackage{graphicx}
\usepackage{float}
\usepackage{caption}
\usepackage{amsthm}
\usepackage{color}
\definecolor{green}{rgb}{0,0.5977,0}
\usepackage[linesnumbered,algoruled,boxed,lined]{algorithm2e}

\newcommand{\rr}{\mathbb R}

\newcommand{\zz}{\mathbb Z}

\newcommand{\ee}{\mathbb E}

\newcommand{\inv}{^{-1}}
\newcommand{\abs}[1]{\left|{#1}\right|}

\newcommand{\suchthat}{\ | \ }

\newcommand{\genseq}[3]{{#1}_1 {#3} {#1}_2 {#3} \dots {#3} {#1}_{#2}}
\newcommand{\seq}[2]{\genseq{#1}{#2}{,}}

\DeclareMathOperator*{\argmax}{argmax}

\newcommand{\txt}[1]{\text{#1}}

\newcommand{\stext}[1]{\ \ \ \ \ \text{(#1)}}
\newcommand{\stextn}[1]{\\&\ \ \ \ \ \ \stext{#1}}

\newcommand{\push}{\\ & \ \ \ \ \ \ \ \ \ \ }

\newcommand{\ipnc}[3]{\begin{figure}[H]\begin{center}\includegraphics[scale = {#1}]{#2.pdf}\caption{#3}\end{center}\end{figure}}

\makeatletter 
\g@addto@macro{\@algocf@init}{\SetKwInOut{Parameter}{Parameters}} 
\makeatother

\newcommand{\lp}[3]{\begin{tabular}{l l}\textbf{#1} & \begin{tabular}{l}$#2$\end{tabular}\\\textbf{subject to} & \begin{tabular}{l l}#3\end{tabular}\end{tabular}}
\newcommand{\lpmax}[2]{\lp{maximize}{#1}{#2}}

\theoremstyle{plain}
\newtheorem{theorem}{Theorem}
\newtheorem{lemma}[theorem]{Lemma} 
\newtheorem{proposition}[theorem]{Proposition}

\newtheorem{conjecture}[theorem]{Conjecture}

\theoremstyle{definition}

\numberwithin{theorem}{section}

\usepackage{url}
\usepackage[hidelinks]{hyperref}
\usepackage[utf8]{inputenc}
\usepackage{graphicx}
\usepackage{amsmath}
\usepackage{amsthm}
\usepackage{booktabs}
\usepackage{float}
\usepackage{graphicx}
\usepackage{caption}
\usepackage{subcaption}
\usepackage{nicefrac}
\usepackage{xcolor}
\usepackage{cleveref}
\usepackage{tagging}
\usepackage{natbib}

\allowdisplaybreaks

\title{Can Buyers Reveal for a Better Deal?}

\author{Daniel Halpern\\ Harvard University \\ \texttt{dhalpern@g.harvard.edu}
\and Gregory Kehne \\ Harvard University \\ \texttt{gkehne@g.harvard.edu} \and
Jamie Tucker-Foltz \\ Harvard University \\ \texttt{jtuckerfoltz@gmail.com}}

\date{}

\begin{document}
\maketitle

\begin{abstract}
    We study market interactions in which buyers are allowed to credibly reveal partial information about their types to the seller. Previous recent work has studied the special case of one buyer and one good, showing that such communication can simultaneously improve social welfare and ex ante buyer utility. However, with multiple buyers, we find that the buyer-optimal signalling schemes from the one-buyer case are actually harmful to buyer welfare. Moreover, we prove several impossibility results showing that, with either multiple i.i.d.\@ buyers or multiple i.i.d.\@ goods, maximizing buyer utility can be at odds with social efficiency, which is surprising in contrast with the one-buyer, one-good case. Finally, we investigate the computational tractability of implementing desirable equilibrium outcomes. We find that, even with one buyer and one good, optimizing buyer utility is generally NP-hard but tractable in a practical restricted setting.
\end{abstract}


\section{Introduction}\label{secIntro}

It is well known that Bayesian-optimal mechanisms for revenue maximization may lead to inefficient outcomes. A seller may rationally refuse to sell to buyers unwilling to pay a high price, even if there is an acceptable lower price at which the seller could still make a substantial profit. But what if a buyer is able to \emph{prove} to the seller that they are unwilling to pay the high price? Upon receiving such a proof, the only rational course of action is for the seller to offer a lower price. As a result, both the buyer and the seller will see a welfare improvement.

However, the possibility of such communication will undoubtedly give rise to secondary market effects. Will the seller infer that a buyer has a higher valuation simply because they do not choose to disclose such a proof, and if so, should the seller raise their price even higher? And if there are multiple buyers competing for a single item, how will the disclosures of one buyer affect the ultimate welfare of another? To realistically discuss the overall welfare implications, it is thus necessary to investigate not just specific possible interactions, but the equilibria of the game played between the seller and the buyer(s).

This inquiry is inspired by the realm of online commerce. The increasing accessibility and quality of buyer data have made the personalized pricing of goods an ever more attractive prospect, and has served as the motivation of previous work studying the impact of information signalling on buyer welfare in auctions~\citep{bergemann15,ali20voluntary}. Motivated by the prospect of a future in which consumers are able to exert precise control over their online data (and a perhaps more immediate future in which sellers implement personalized pricing), we aim to answer the question,

\begin{quote}
    \emph{``Can consumers benefit from the ability to share their private data, and if so, how?''}
\end{quote}

In recent work, \citet*{ali20voluntary} initiate the study of how such \emph{voluntary disclosure} capabilities can improve welfare, considering a handful of special cases. In their model, a prospective buyer is allowed to credibly disclose to the seller a set of possible types containing their true type; the seller then sets prices based on this information. They report overwhelmingly positive news for consumers. When there is one buyer, one seller, and one good, they demonstrate that there always exists a disclosure strategy for the buyer such that
\begin{itemize}
    \item the buyer has no incentive to deviate from the strategy after learning their type (technically, the strategies are part of a sequentially-rational Bayes-Nash equilibrium),
    \item the good is always sold,
    \item the seller is weakly better off than they would be without disclosure, and
    \item every interim buyer type is weakly better off than they would be without disclosure.
\end{itemize}
For a parameterized family of canonical probability distributions over the buyer's value for the good (including the uniform distribution on $[0, 1]$), they show that it is possible to strictly increase ex ante buyer utility as well. Furthermore, there is an intuitive characterization of the buyer-optimal equilibrium, determined by the limit of a greedy algorithm that iteratively constructs better and better equilibria by having all buyer types who are not sold the good declare to the seller that they are of such a type. In the end, we are left with a \textit{partitional equilibrium}, in which there is some partition $\mathcal{P}$ of the types and every buyer reveals the set in $\mathcal{P}$ to which their type belongs.

\tagged{DeleteThisTagForArxivVersion}{\citet*{ali20voluntary} also study a case of multiple sellers; in particular, a setting in which there are two sellers, and the lone buyer has strong but private preferences over from whom to buy. Here they show that the buyer may leverage selective (seller-personalized) disclosure in order to play sellers off of one another and again compel their personalized prices to increase the buyer's own expected utility.}

However, the settings of these results differ markedly from most online commerce, and it is in the direction of these differences we depart.

In \Cref{secMultipleUniform01Buyers} we investigate the effects of disclosure when there are \emph{two} i.i.d.\@, uniform $[0, 1]$ buyers instead of one. Surprisingly, we find that the natural, buyer-symmetric analogues of the optimal one-buyer equilibria from \citet*{ali20voluntary} no longer yield buyer welfare improvements. Perhaps even more surprisingly, it is possible to improve the expected buyer surplus (the sum of the buyers' utilities) by having only one buyer disclose information about their type (though this harms the utility of the other buyer). As for the question of social efficiency, with a few additional assumptions in the spirit of~\citep{ali20voluntary}, we show an extreme impossibility result (\Cref{thmNoTwoBuyerEfficiency}): in any equilibrium where the good is always allocated to the highest bidder, both buyers must always receive utility zero. Since the model assumes the seller has no cost to sell the goods, this shows that social efficiency is incompatible with maximizing buyer welfare, which lies in stark contrast to the one-buyer, one-good case.

In \Cref{secGeneralImpossibility} we further generalize these impossibility results to settings with richer disclosure capabilities and arbitrary priors. We show that, with either multiple buyers \emph{or} multiple goods, maximizing buyer surplus may require the seller to sometimes \emph{not} sell all of the goods (\Cref{thmGeneralImpossibility}). This holds even with the restrictions that buyer valuations are additive and independent across goods, as well as independent across buyers.

Finally, in \Cref{secComplexity} we study the problem of maximizing consumer welfare through disclosure schemes from a computational perspective. We model this problem by approximating arbitrary priors by discrete probability distributions with finite support, which are encoded as part of the input. We show that, while it is possible to efficiently compute the buyer-optimal equilibrium in the restricted setting from~\citep{ali20voluntary} where disclosure messages must be ``connected'' (\Cref{thmDP}), the more general problem is (weakly) NP-hard (\Cref{thmHardness}), and is inapproximable by connected equilibria (\Cref{proNoApprox}).

\subsection{Related work}

This work falls within a larger body of literature on the implications of information signalling in markets and how strategic disclosure affects equilibria, as in the work of \citet*{gentzkow17bayesian} on \textit{Bayesian persuasion}. On a more conceptual level, it contributes to a growing literature on the connection between privacy and information in markets. See \citet*{acquisti2016economics} and \citet*{bergemann2019markets} for relevant surveys.

The signals which we consider are the \textit{verifiable disclosures} of~\citep{ali20voluntary}, in which buyers send public signals about their types from a set of possible signals which are demonstrably truthful. The study of verifiable disclosure dates back to classic economic works by \citet*{grossman1981informational} and \citet*{milgrom1981good}, and has recently seen a resurgence in the economics literature, both in market settings and more general games~\citep{kartik2012implementation,ben2012implementation,hagenbach2014certifiable,hart2017evidence,ben2018disclosure,ben2019mechanisms,koessler2019selling}. The original game focused on a seller being able to verifiably disclose information about the item they are selling, rather than the buyer about their type. A closely related work by \citet*{sher15discrimination} considers the setting where the disclosures are verifiable, and there is commitment from both the buyer and the seller.


\ipnc{.55}{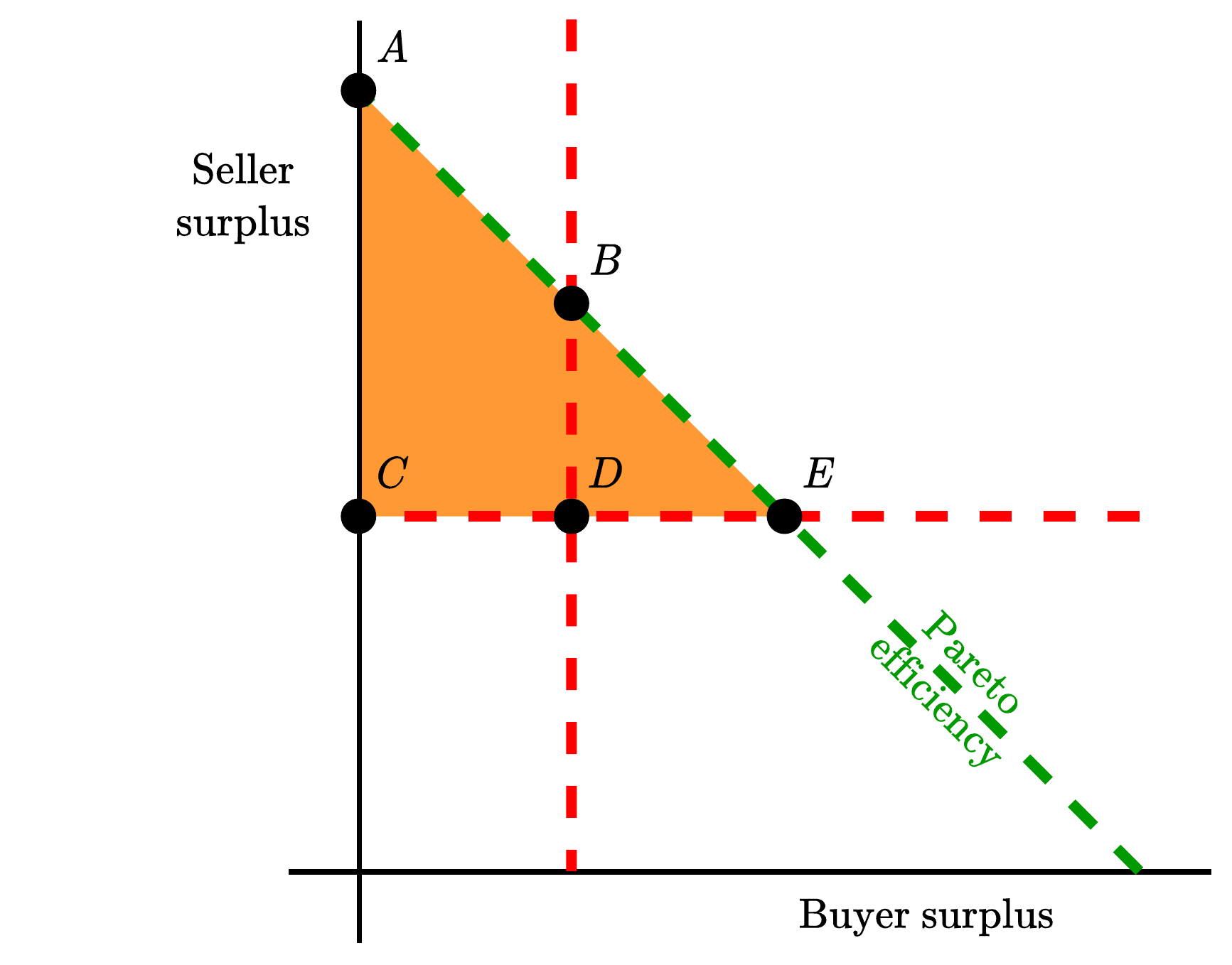}{Feasible welfare outcomes from \citet*{bergemann15}
\label{figBBMTriangle}}

In terms of \Cref{figBBMTriangle}, in our setting \emph{with} verifiability but \emph{without} commitment to a signalling scheme, the main result of \citet*{ali20voluntary} is that, if point $D$ in \Cref{figBBMTriangle} represents the welfare of the buyer and seller in the absence of any communication by the intermediary, then there is some feasible point on line segment $BE$ when there is only one buyer and one good. We show that this is false for either multiple buyers or multiple goods. In fact, for multiple goods, we give an example where the only feasible point on segment $AE$ is at $A$, where the buyer receives no surplus at all (see \Cref{secGeneralImpossibility}).

A parallel line of work aims to improve buyer welfare via an intermediary that observes buyer valuations and then sends a signal to the seller based on some prearranged signalling scheme. \citet*{bergemann15} address the case of one buyer and one seller. They show that any combination of welfares in the shaded triangle is attainable via randomized signalling schemes. \citet*{IntermediaryMultipleBuyers} generalize this to the setting with multiple buyers who share an intermediary, showing that the line segment $BE$ is unattainable, though it is possible to approximate the maximum possible buyer welfare when the buyers' types are independent and identically distributed.

Similar to our setting is that of \citet*{shen19signaling}, which generalizes \citet*{bergemann15} to the case of multiple buyers. For them, the buyers each simultaneously signal according to pre-committed signalling schemes to the seller, who then conducts a Myerson auction based on their updated priors.
They study the equilibria of this game and determine buyer-optimal equilibria for certain classes of buyer type distributions, but the buyer disclosures are neither deterministic nor verifiable. This setting also differs from ours in that buyer signalling schemes are required to be best responses to each other, a constraint that we do not enforce.


\section{Model}\label{secModel}

As in~\citep{ali20voluntary}, we operate in the context of a verifiable \emph{disclosure game}. We begin by defining the most general, abstract form of the game, with multiple buyers and multiple goods but only one seller. Suppose there are $m$ goods, numbered $1, 2, \dots, m$. We are concerned only with additive valuations, so we denote the type space of each buyer by $\rr_{\geq 0}^m$, where, for any $v = (v_1, v_2, \dots, v_m) \in \rr_{\geq 0}^m$, each $v_k$ denotes the value the buyer has for good $k$. There is a common prior over the buyers' values, in which the value any fixed buyer has for different goods may be correlated; however, in all of the multiple-buyer scenarios we consider in this paper, the values of different buyers are independent. The disclosure game proceeds in two stages:
\begin{enumerate}
    \item\label{itmGameMessage} Each buyer simultaneously observes their value ${v \in \rr_{\geq 0}^m}$ and publicly sends a \emph{message} in the form of a set ${M \subseteq \rr_{\geq 0}^m}$ such that ${v \in M}$.
    \item\label{itmGameSell} The seller sells the good(s) to the buyer(s) so as to maximize expected revenue, taking into account the information conferred by $M$.
\end{enumerate}
Note that the requirement that $v \in M$ is a key feature of this game: buyers cannot misrepresent their types in the disclosure stage. We study the subgame-perfect pure-strategy Bayes-Nash equilibria of this game and evaluate them with respect to \emph{ex ante buyer surplus}, defined as the expected sum of all buyer utilities.

With one buyer and one good, step~\ref{itmGameSell} simply involves the seller choosing a posted price and the buyer accepting or rejecting. With one buyer and multiple goods, the seller posts a menu of bundles of goods, each with an associated price, and the buyer may choose one of them. With one good and multiple buyers, the seller runs a Myerson auction.\footnote{See \citep{myerson1981optimal} for a description and discussion of virtual values which we use extensively. In this case, step~\ref{itmGameSell} implicitly involves additional communication from the buyers to the seller, revealing their types as prescribed by the direct revelation mechanism. This communication is different from that of step~\ref{itmGameMessage} in that now buyers are allowed to lie about their types.} We introduce new notation to describe these specialized settings as needed.

\subsection{Special equilibria}\label{subEfficiencyAndPartitions}

An equilibrium is \emph{efficient} if the good(s) are always sold to the buyer with the highest value. An equilibrium is \emph{partitional} if each buyer's messaging strategy is induced by a partition $\mathcal{P}$ of $\rr_{\geq 0}^m$, where the buyer reveals to which set in $\mathcal{P}$ their value belongs. A \emph{connected} partitional equilibrium is one in which the interiors of the convex hulls of the messages in $\mathcal{P}$ are disjoint, in which case we say the messages are \emph{connected} and that $\mathcal{P}$ is a \emph{connected partition}. As it is shown in~\citep{ali20voluntary}, to maximize expected buyer utility in the one-buyer one-good case, it is without loss of generality to restrict attention to efficient, partitional equilibria.

\begin{lemma}[Efficiency Lemma \citep{ali20voluntary}] \label{lemEfficiency}
    Suppose there is one good and one buyer. Given any pure-strategy equilibrium of the disclosure game, there exists a pure-strategy equilibrium that is efficient and results in the same payoff for every buyer type.
\end{lemma}

\begin{lemma}[Partitional Lemma \citep{ali20voluntary}] \label{lemPartitional}
    Given any pure-strategy equilibrium of the disclosure game, there exists a pure-strategy equilibrium that that is partitional and results in the same selling mechanism for every buyer type.
\end{lemma}

The Efficiency Lemma is proved by having all buyer types that do not get the good reveal their type to the seller. One can easily check that this does not change seller incentives when faced with one of the other buyer types. The Partitional Lemma is proved by partitioning the buyer types by the message that each type sent in the original equilibrium, thus conveying the same information to the seller.\footnote{The version of the Partitional Lemma proved in~\citep{ali20voluntary} is slightly different since it is proved in the restricted setting where seller messages are required to be connected intervals. It only applies to efficient equilibria and only guarantees the same selling mechanism for \emph{almost} every buyer type.} While the partitional lemma continues to hold for multiple buyers and/or multiple goods, we show in \Cref{secGeneralImpossibility} that the Efficiency Lemma does not.


\section{Disclosure with two uniform [0, 1] buyers}\label{secMultipleUniform01Buyers}

For any $a \leq b$, let $U[a, b]$ denote the uniform distribution on $[a, b]$. In this section we consider the  canonical setting where there are two buyers, $A$ and $B$, with valuations $v_A$ and $v_B$ for a single good drawn independently from $U[0, 1]$.

By the Partitional Lemma (\Cref{lemPartitional}), we may restrict attention to equilibria of the disclosure game for which $A$ and $B$ report messages $P_A$ and $P_B$ from $\mathcal{P}_A$ and $\mathcal{P}_B$, which are partitions of $[0, 1]$. In this section, we will only be concerned with the special case where $\mathcal{P}_A$ and $\mathcal{P}_B$ are connected, i.e., each element is an interval. Again, as in the one-buyer setting, all pairs of partitions $\mathcal{P}_A, \mathcal{P}_B$ are supportable as equilibria in this game. If the seller holds the off-path belief that, upon receiving from $A$ any message $M \not \in \mathcal{P}_A$, the valuation of $A$ is $\overline{v_A} := \max\{M\}$ with probability 1,\footnote{This maximum may not be well-defined in general, but it is well-defined for connected, partitional equilibria with a $U[0, 1]$ prior, since we may ensure all messages are closed on the top end. This can only improve buyer utilities, and does not change seller incentives since $U[0, 1]$ has no atomic points.} then $A$ is guaranteed not to derive any utility from the resulting Myerson auction, since in order to receive the good they must clear the Myerson reserve of $\overline{v_A} \geq v_A$. 
Facing a seller with these off-path beliefs, $A$ is therefore incentivized to report the unique $P_A \in \mathcal{P}_A$ for which $v_A \in P_A$, since otherwise $A$ is guaranteed to receive no utility.
The same argument, of course, applies to $B$.

Throughout this section, we make repeated use of the following calculation.

\begin{lemma}\label{lemUniformvv}
    A buyer with value $v \sim U[a,b]$ has virtual value $\phi(v) = 2v - b$.
\end{lemma}
\begin{proof}
    For this distribution, $F(v) = \frac{v-a}{b-a}$ and $f(v) = \frac{1}{b-a}$, and so $\phi(v) = v - \frac{1 - F(v)}{f(v)} = 2v - b$.
\end{proof}

\subsection{Example: the first step of Zeno's partition}\label{subMultipleBuyersDiabolo}

In~\citep{ali20voluntary}, the optimal equilibrium for the one-buyer $U[0, 1]$ distribution is induced by ``Zeno's partition,"
$$\mathcal{P}_Z := \{(2^{-k-1}, 2^{-k}] \suchthat k \in \zz_{\geq 0}\} \cup \{\{0\}\}.$$
It is constructed through a sequence of steps from the no-disclosure equilibrium. In each step, all buyer types who are currently not sold the good are separated into a new element of the partition. In the one-buyer case, each step is a Pareto improvement for all buyer types; let us now consider the two-buyer case and see what happens when we implement the first step of Zeno's partition for both buyers simultaneously.

\begin{figure*}
  \begin{subfigure}{0.24\textwidth}
    \includegraphics[width=\textwidth]{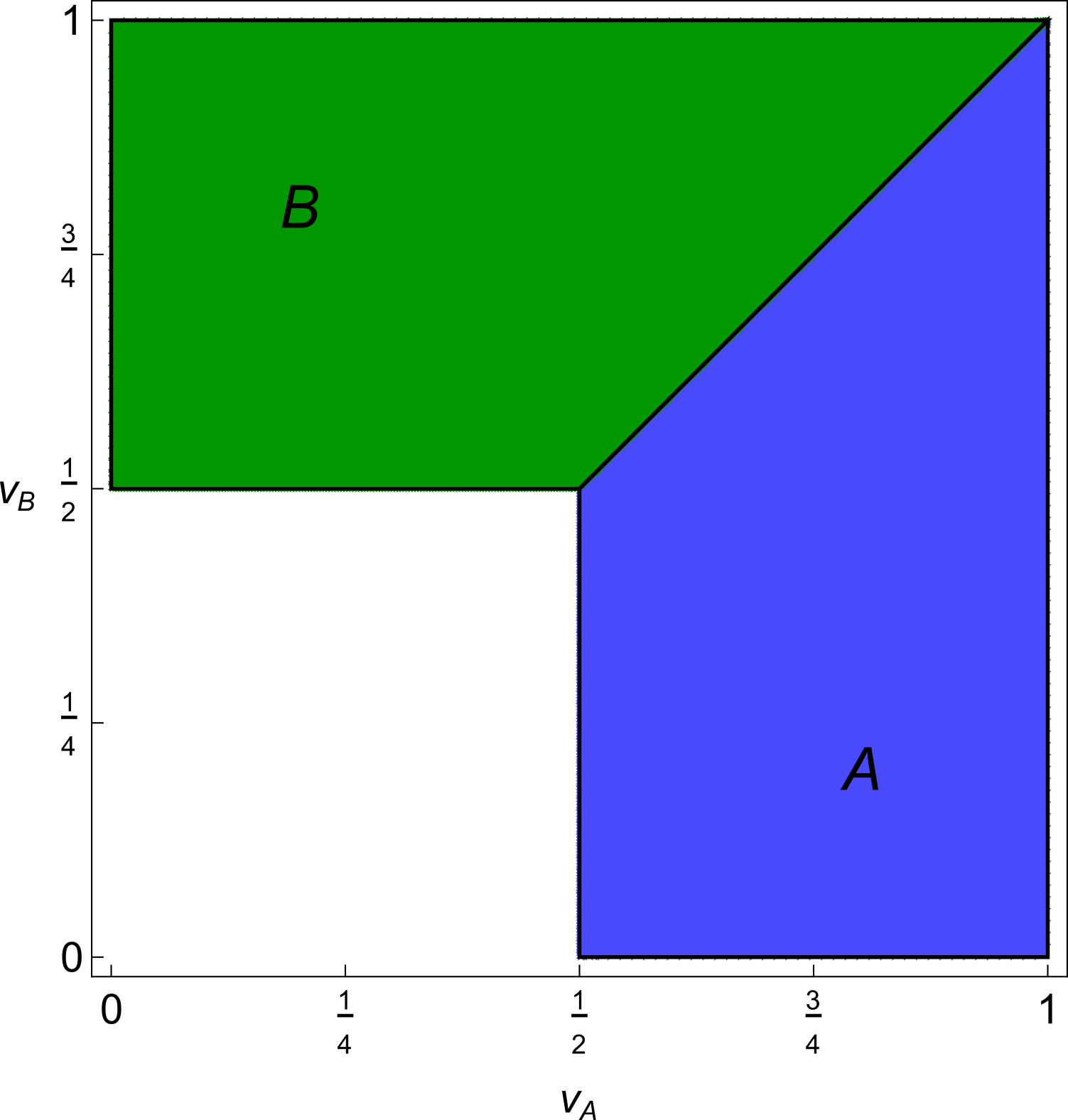}
    \caption{No disclosure: $\mathcal{P}_A = \mathcal{P}_B = \{[0,1]\}$, yielding an expected buyer surplus of $\mathbb{E}[U]=\frac{1}{6}$.}
    \label{subfig:nodisclose}
  \end{subfigure}%
  \hspace*{\fill}   
  \begin{subfigure}{0.24\textwidth}
    \includegraphics[width=\textwidth]{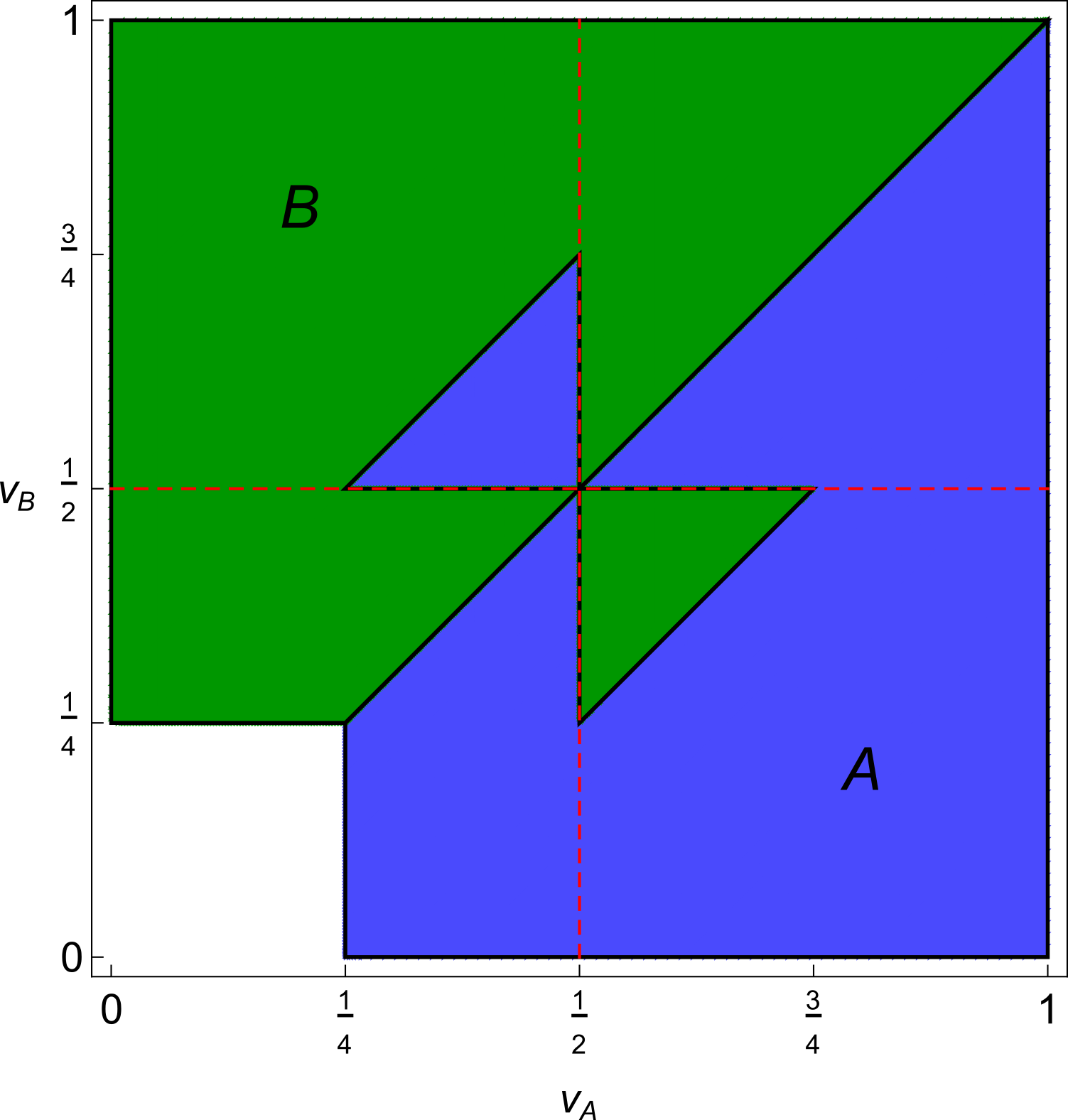}
    \caption{High/low disclosure: $\mathcal{P}_A = \mathcal{P}_B = \{[0,1/2],[1/2, 1]\}$. $\mathbb{E}[U]=\frac{1}{6}$.}
    \label{subfig:diabolo}
  \end{subfigure}
  \hspace*{\fill}   
  \begin{subfigure}{0.24\textwidth}
    \includegraphics[width=\textwidth]{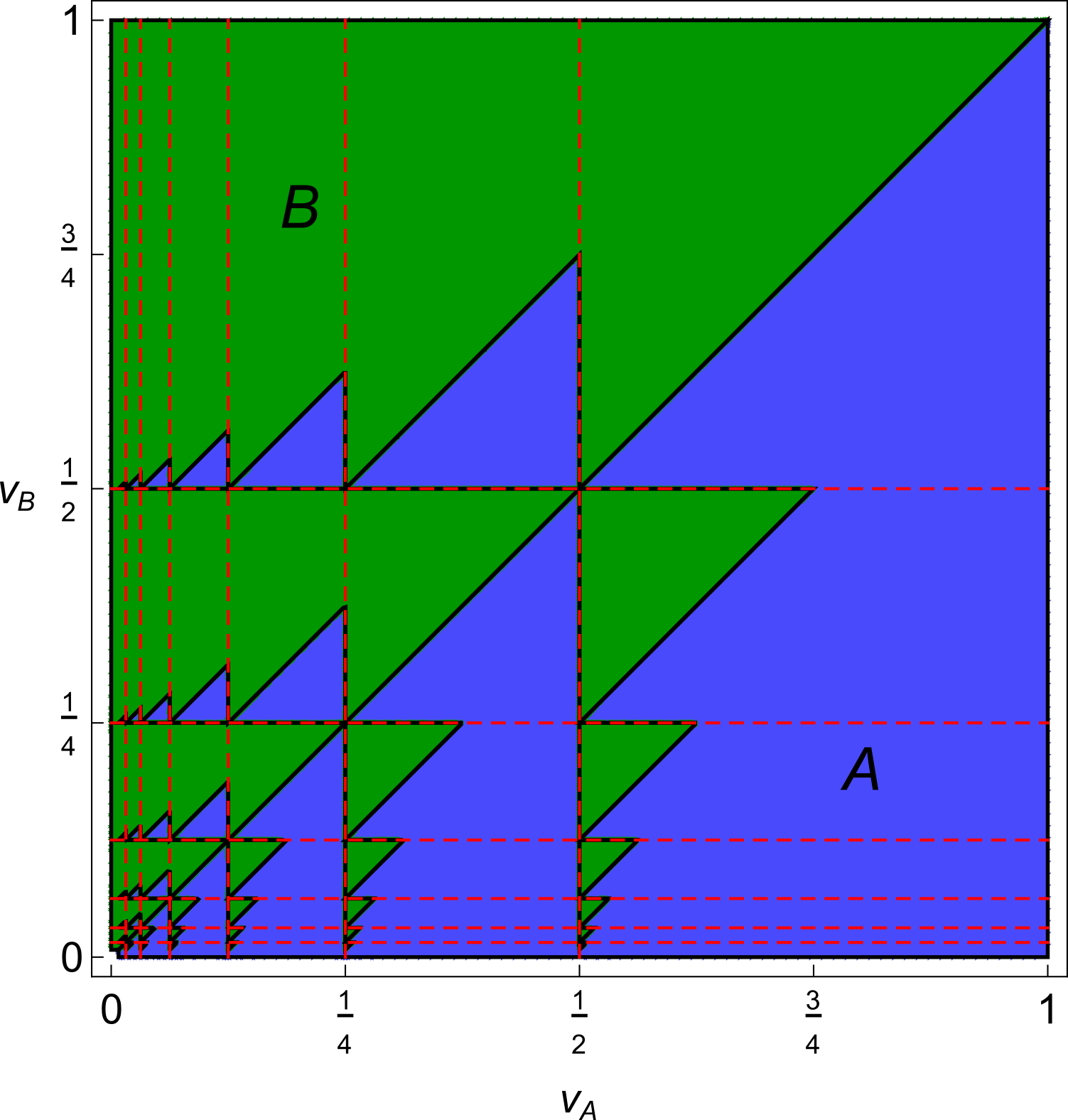}
    \caption{$A$ and $B$ disclose according to Zeno's partition $\mathcal{P}_Z$.\\ $\mathbb{E}[U]=\frac{23}{147}<\frac{1}{6}$.}
    \label{subfig:1d}
  \end{subfigure}%
  \hspace*{\fill}   
  \begin{subfigure}{0.24\textwidth}
    \includegraphics[width=\textwidth]{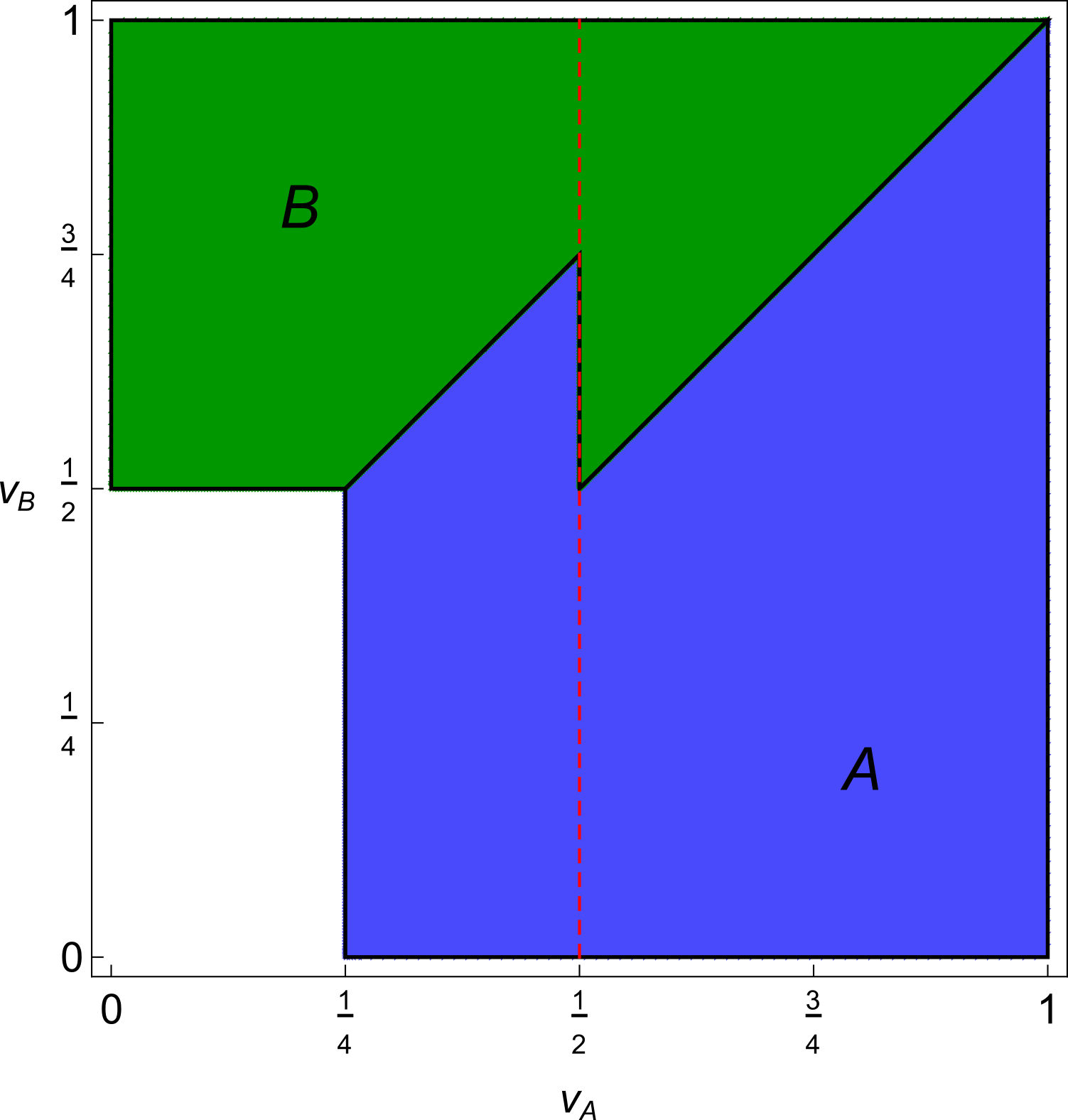}
    \caption{Coarse asymmetric disclosure: $\mathcal{P}_A = \{[0,\frac{1}{2}],$ $[\frac{1}{2},1]\}$ and $\mathcal{P}_B = \{[0,1]\}$. $\mathbb{E}[U]=\frac{11}{64} > \frac16$.} \label{subfig:1b}
  \end{subfigure}%
\caption{Allocations of the good in different partitional equilibria. Blue denotes that $A$ receives the good, green denotes that $B$ receives the good, and white denotes no sale. Expected buyer surplus $\mathbb{E}[U]$ is given for each case, and the dashed red line indicates the boundary between parts in the partitions.}
\label{fig:experimentallocations}
\end{figure*}

In the no-disclosure equilibrium (where each buyer always sends the message $[0, 1]$), the seller runs a second-price auction with reserve price $\frac12$. A simple calculation shows that the expected buyer surplus is $\frac16$. For a plot of the allocation as a function of $(v_A, v_B)$ see \Cref{subfig:nodisclose}.

Now consider what happens when buyers reveal whether their valuation is greater than or less than $\frac12$. There are 3 cases to consider. If both buyers have value less than $\frac12$, then the seller will run a second-price auction with reserve price $\frac14$, so the result will be the same as in the no-disclosure case, but with everything scaled down by a factor of 2. Thus, the expected buyer surplus will be $\frac12 \cdot \frac16 = \frac{1}{12}$. If both buyers have value $\geq \frac12$, it is not to hard to check that the seller still runs a second-price auction with a reserve price of $\frac12$ (or equivalently, no reserve price). Thus we have the same outcome as in the no-disclosure case for an expected buyer surplus of $\frac16$.

So far, in the first two cases, everything is analogous to the situation with one buyer: if the low-value types disclose that they have low value, we see an improvement in buyer welfare, whereas if there is no disclosure, the seller incentives remain the same, so the buyers achieve the same welfare. However, in the third case, where one buyer discloses they have a low value and the other does not, something quite different happens: we see competition between the two buyers that ultimately reduces the welfare of the high-value buyer.

The sale outcomes in the partitional equilibrium are depicted in \Cref{subfig:diabolo}. The case where buyer $A$ discloses ``low'' and buyer $B$ discloses ``high'' corresponds to the top-left quadrant. The small blue triangle represents the case where the seller sells the good to $A$ because, even though they have a lower value, they have a \emph{higher virtual value}.

We can immediately deduce that, for small $\varepsilon > 0$, a buyer with value $\frac12 + \varepsilon$ gets a lower interim expected utility compared to the no-disclosure equilibrium. For in the no-disclosure equilibrium, they are sold the good with probability roughly $\frac12$, and when they do win, they gain an expected utility of roughly $\varepsilon$. However, in this new equilibrium, they are only sold the good with probability roughly $\frac14$, yet still only gain an expected utility of roughly $\varepsilon$ when they win. Thus, in contrast to the one buyer case, the new equilibrium is not a Pareto improvement across buyer types. In fact, we will see that it gives \emph{exactly the same} ex ante utility for the buyers.

Suppose that buyer $A$ reveals that $v_A \in [0, \frac12]$ and buyer $B$ reveals that $v_B \in [\frac12, 1]$. (The other two cases are addressed above.) Applying \Cref{lemUniformvv}, the virtual values of the two buyers and their inverses are as follows:
\begin{align*}
	\phi_A(v_A) &= 2v_A - \frac12 & \phi_B(v_B) &= 2v_B - 1 & 
	\phi_A\inv(x) &= \frac{x}{2} + \frac14 & \phi_B\inv(x) &= \frac{x}{2} + \frac12.
\end{align*}
Note that, for $v_B \in [\frac12, 1]$, $\phi_B(v_B) \geq 0$, so the good is always sold under the Myerson auction. However, for $v_B < v_A + \frac14$, the good will be sold to buyer $A$ since $\phi_B(v_B) < \phi_A(v_A)$, even though they have a lower valuation.

To compute the expected buyer surplus in the case where $v_A \in [0, \frac12]$ and $v_B \in [\frac12, 1]$, we break our computation up into 3 disjoint sub-cases:
\begin{itemize}
	\item $A$ has value less than $1/4$ so $B$ wins with utility $v_B - \phi_B\inv(0)$.
	\item $A$ has value greater than $1/4$, but $\phi_B(v_B) \geq \phi_A(v_A)$, so $B$ still wins with utility $v_B - \phi_B\inv(\phi_A(v_A))$.
	\item $\phi_B(v_B) < \phi_A(v_A)$, so $A$ wins with utility $v_A - \phi_A\inv(\phi_B(v_B))$.
\end{itemize}
We compute the expected buyer surplus by computing the corresponding 3 integrals with Mathematica. The expected buyer surplus from the case where $v_A \in [0, \frac12]$ and $v_B \in [\frac12, 1]$ is
\begin{align*}
	\ee[U] &= \int_{0}^{1/4}\int_{1/2}^{1} (v_B - \phi_B\inv(0)) \, dv_B dv_A \push + \int_{1/4}^{1/2}\int_{v_A + 1/4}^{1} (v_B - \phi_B\inv(\phi_A(v_A))) \, dv_B dv_A \push + \int_{1/4}^{1/2}\int_{1/2}^{v_A + 1/4} (v_A - \phi_A\inv(\phi_B(v_B))) \, dv_B dv_A = \frac{5}{96}.
\end{align*}
The case where $v_B \in [0, \frac12]$ and $v_A \in [\frac12, 1]$ is symmetric, so the expected buyer surplus coming this case is also $\frac{5}{96}$.

Adding up the expected buyer surplus in each of the four cases, we have that the overall expected buyer surplus is $$\frac14 \cdot \frac{1}{12} + \frac14 \cdot \frac16 + 2 \cdot \frac{5}{96} = \frac{1}{48} + \frac{2}{48} + \frac{5}{48} = \frac{8}{48} = \frac16.$$

Thus, we have the same expected buyer surplus, $\frac16$, as in the no-disclosure equilibrium. All we have accomplished is a redistribution of welfare from higher types to lower types, and we have introduced new inefficiencies.

\subsection{The search for better equilibria}\label{subMultipleGoodsBuyersPartitions}

In an effort to better understand the equilibria of this disclosure game when $v_A, v_B \sim U[0,1]$, we conducted a computer search over the space of partitional equilibria $\mathcal{P}_A$, $\mathcal{P}_B$ which partition $[0,1]$ into (a reasonable number of) intervals.

We begin with the observation that, for given $v_B$, the expectation of the payment $p_A$ charged to $A$ is 
\[
    \mathop\mathbb{E}_{v_A \sim U[a,b]}[p_A | v_B] = \mathbb{E}_{v_A}[p_A] = \mathbb{E}_{v_A}[x(v_A, v_B) \cdot \phi_A(v_A))],
\]
and so the expected payoff $u_A$ of $A$ over $v_A \sim U[a,b]$ and $v_B \sim U[c,d]$ is
\[
    \mathop\mathbb{E}_{v_B \sim U[c,d]} \mathop\mathbb{E}_{v_A \sim U[a,b]}[u_A] = \mathop\mathbb{E}_{v_B \sim U[c,d]} \mathop\mathbb{E}_{v_A \sim U[a,b]}[x(v_A, v_B) \cdot (v_A - \phi_A(v_A))].
\]
Applying \Cref{lemUniformvv} to $v_A \sim U[a,b]$ and simplifying then lets us write the expected payoff of buyer $A$ as
\begin{equation} \label{eq:buyersurplussimplification}
    \mathop\mathbb{E}_{v_B \sim U[c,d]} \mathop\mathbb{E}_{v_A \sim U[a,b]}[u_A] = \frac{1}{(b-a)(d-c)} \int_R b - v_A \: dv_A \: dv_B,
\end{equation}
where $R$ is the region of $(v_A, v_B)$ given by the inequalities
\begin{align*}
    a \leq v_A &\leq b, &
    c \leq v_B &\leq d, &
    v_A &\geq b/2, &
    v_A &\geq v_B + \frac{b-d}{2}.
\end{align*}
Finally, the buyer surplus is just $U = u_A + u_B$, so for a given region it is simply
\[
     \mathop\mathbb{E}_{v_B \sim U[c,d]} \mathop\mathbb{E}_{v_A \sim U[a,b]}[U] =  \mathop\mathbb{E}_{v_B \sim U[c,d]} \mathop\mathbb{E}_{v_A \sim U[a,b]}[u_A] +  \mathop\mathbb{E}_{v_B \sim U[c,d]} \mathop\mathbb{E}_{v_A \sim U[a,b]}[u_B].
\]

Once the boundaries of $R$ are determined, this integral can be computed symbolically.
Therefore computing the expected buyer surplus for a given pair of partitions $\mathcal{P}_A$, $\mathcal{P}_B$ simply entails evaluating the above expression for each pair of possible intervals $(P_A, P_B)$.

Our computational results support the following conjecture:
\begin{conjecture}
    For $v_A, v_B \sim U[0,1]$ the partitional equilibrium given by $\mathcal{P}_A = \{[0,\frac{1}{2}], [\frac{1}{2}, 1]\}$ and $\mathcal{P}_B = \{[0, 1]\}$ maximizes expected buyer welfare for the class of partitional equilibria.
\end{conjecture}

This disclosure profile (with allocations shown in \Cref{subfig:1b}) is notable in that it is asymmetric, and in that it appears to strictly outperform all symmetric profiles in terms of expected buyer surplus.
But most surprisingly, the buyer who discloses (in this case, buyer $A$) has increased expected utility, while the buyer who does not disclose suffers a strict decrease in expected utility as compared to the symmetric no-disclosure equilibrium (\Cref{subfig:nodisclose}); $A$ receives expected utility $\frac{13}{128} \approx 0.102$, while $B$ receives $\frac{9}{128} \approx 0.070$. The total expected surplus is $\frac{11}{64} > \frac16$.

At the same time, in this asymmetric regime there is a tradeoff between disclosure detail and likelihood of receiving the good on the one hand, and expected surplus on the other.
But as $A$ discloses more information and $B$ does not disclose, $A$'s expected utility and relative surplus begins to decrease.
In the limit, where $A$ discloses $v_A$ exactly and $B$ discloses nothing, their respective expected utilities tend towards $\ee[U_A] = 0$ and $\ee[U_B] = 1/24$.

Experimental evidence also suggests that:

\begin{conjecture}
    For $v_A, v_B \sim U[0,1]$ nondisclosure maximizes expected buyer welfare for the class of all symmetric message sets which are interval partitions.
\end{conjecture}

This no-disclosure equilibrium yields an expected buyer surplus of $\frac{1}{6}$, the same as the equilibrium analyzed in \Cref{subMultipleBuyersDiabolo}.
In fact, \textit{all} symmetric buyer disclosure profiles with a single high/low threshold $t \in [0,\frac{1}{2}]$ realize an expected buyer welfare of $\frac{1}{6}$. (See \Cref{appOneSixth} for a proof.)

In contrast to the one-buyer case, the equilibrium with symmetric disclosure given by Zeno's partition (\Cref{subfig:1d}) confers expected buyer utility $\frac{23}{147} < \frac{1}{6}$.

\subsection{Efficiency versus buyer welfare}

If both buyers fully disclose their types, then the seller will sell to the buyer with a higher value at that value, so we will have a socially efficient outcome, but neither buyer will receive any utility. On the other hand, in all of the partial disclosure equilibria considered thus far, including no-disclosure, some buyer gets nonzero utility, but the outcome is not guaranteed to be efficient (the good may be sold to the buyer with a lower value, or not sold at all). A natural question is whether it is possible to simultaneously achieve efficiency and nonzero buyer surplus. Our first main result is that, if we restrict attention to connected partitional equilibria, the answer is negative.

\begin{theorem}\label{thmNoTwoBuyerEfficiency}
    For $v_A, v_B \sim U[0,1]$, there are no partitions $\mathcal{P}_A,\mathcal{P}_B$ of $[0,1]$ into intervals inducing an efficient equilibrium that confers nonzero expected buyer surplus.
\end{theorem}

\begin{proof}
    We assume that the equilibrium induced by $\mathcal{P}_A$ and $\mathcal{P}_B$ confers nonzero buyer surplus, and show that it must be inefficient for some $v_A, v_B \in [0, 1]$. 
    The only property we actually need to assume is that there exists some interval $P_A \in \mathcal{P}_A$ of measure greater than zero. 
    This follows without loss of generality, for if all intervals in both partitions have measure zero, i.e., are singletons, then both buyers always exactly disclose their values, and hence get zero utility. 
    Let $0 \leq a < b \leq 1$ be such that the closure of $P_A$ is $\overline{P_A} = [a, b]$. 
    Let
    $$x := \frac{\max\{0, 2a - b\} + a}{2}.$$
    Note that $x$ is the average of $a$ and some point weakly less than $a$, so $x \leq a$ (and $x = 0$ precisely when $a = 0$; otherwise $x$ is in the open interval $(0, a)$). 
    Let $P_B$ be the unique element of $\mathcal{P}_B$ containing $x$, and let $0 \leq c \leq d \leq 1$ be such that $\overline{P_B} = [c, d]$. 
    There are five cases to consider.
    
    \textbf{Case 1:} $x = 0$. 
    In this case, it must be that $a = c = 0$. Let $v_A := \frac{b}{4} \in P_A$ and $v_{B} := 0$. 
    After disclosing that their type is in $[a, b]$, by \Cref{lemUniformvv}, the virtual value of buyer $A$ is
    $$\phi_A = 2v_A - b = -\frac{b}{2} < 0,$$
    so they will not get the good. 
    This is inefficient since $v_A > v_B = 0$.
    
    \textbf{Case 2:} $x > 0$ and $d \leq a$. For small $\varepsilon > 0$, let $v_A := a + \varepsilon \in P_A$ and $v_B := \frac{x + d}{2} \in P_B$. Then by \Cref{lemUniformvv} the virtual value of buyer $A$ is
    $$\phi_A = 2v_A - b = 2a - b + 2\varepsilon$$
    and the virtual value of buyer $B$ is
    $$\phi_B = 2v_B - d = x \geq \frac{3a - b}{2} > \frac{4a - 2b}{2} = 2a - b = \phi_A - 2\varepsilon.$$
    Therefore, for sufficiently small $\varepsilon$, buyer $B$ will get the good, which is inefficient since
    $$v_B = \frac{x + d}{2} \leq \frac{x + a}{2} \leq \frac{a + a}{2} = a = v_A - \varepsilon < v_A.$$
    
    \textbf{Case 3:} $a < d < b$. For small $\varepsilon \geq 0$, let $v_A := d + \frac{b - d}{4} \in P_A$ and $v_B := d - \varepsilon \in P_B$. 
    (Note that the $\varepsilon$ is needed to ensure we have $v_B \in P_B$. Choosing $\varepsilon > 0$ is necessary if $d \notin P_B$; choosing $\varepsilon = 0$ is necessary if $P_B = \{c\} = \{d\}$.)
    Then the virtual value of buyer $A$ is
    $$\phi_A = 2v_A - b = \frac{3d - b}{2}$$
    and the virtual value of buyer $B$ is
    $$\phi_B = 2v_B - d = d - 2\varepsilon = \frac{2d}{2} - 2\varepsilon > \frac{3d - b}{2} - 2\varepsilon = \phi_A - 2\varepsilon.$$
    Therefore, for sufficiently small $\varepsilon$, buyer $B$ will get the good, which is inefficient since
    $$v_B = d - \varepsilon < d + \frac{b - d}{4} = v_A.$$
    
    \textbf{Case 4:} $d > b$. Observe that $c \leq x \leq a < b$. Thus, we have $c < b < d$, so this is the same as Case 3, with the two roles of the two buyers reversed.
    
    \textbf{Case 5:} $x > 0$ and $d = b$. Let $x'$ be any point in the open interval $(x, a)$. Let $P'_A$ be the unique element of $\mathcal{P}_A$ containing $x'$, and let $0 \leq a' \leq b' \leq a$ be such that $\overline{P'_A} = [a', b']$. Observe that $c \leq x < x' \leq b' \leq a < b = d$. Since $c < b' < d$, we are again in Case 3, with the roles of the two buyers reversed, and substituting $P'_A$ for $P_A$.
\end{proof}


\section{General impossibility results}\label{secGeneralImpossibility}

In this section, we prove our second main result, which, unlike Theorem \ref{thmNoTwoBuyerEfficiency}, holds with respect to \emph{arbitrary} pure-strategy equilibria, not just those induced by connected partitions.

\begin{theorem}\label{thmGeneralImpossibility}
    In any setting with multiple goods or multiple buyers, there exist common priors over buyer valuation functions such that:
    \begin{itemize}
        \item The buyer valuation functions are pairwise independent across different buyers.
        \item Every buyer's valuation function is additive and independent across different goods.
        \item In any pure-strategy equilibrium of the disclosure game in which all goods are always sold, buyer surplus is strictly lower than it would be in the absence of disclosure. (Additionally, in the case of multiple goods, this buyer surplus must be zero.)
    \end{itemize}
\end{theorem}

This implies that the Efficiency Lemma does not hold beyond the limited context of one good and one buyer. In other words, social efficiency may be incompatible with maximizing expected buyer welfare. We begin by discussing the computational approach underlying our proof.

\subsection{Computing the optimal mechanism over discrete distributions}\label{subComputationLP}

Suppose the buyers have sent messages to the seller, and now the seller is deciding how to optimally sell the goods. By the revelation principle, it is without loss of generality to consider selling mechanisms in which each buyer is asked to exactly reveal the values they have for each good (but, unlike in the initial disclosure, now they are allowed to lie, so the mechanism must be incentive compatible). If there are only a finite number of types each buyer can have, then we can write the seller's optimization problem as a linear program, as follows.

Suppose there are $\ell$ buyers and $m$ goods. For any $j \in [\ell] := \{1, 2, \dots, \ell\}$, suppose there are $n_j$ possible types that buyer $j$ can have. Let $T$ denote the joint type space of all buyers,
$$T := \prod_{j \in [\ell]} [n_j],$$
where an element $\mathbf{i} \in T$ denotes a vector of types for each buyer, $\mathbf{i} = \seq{i}{\ell}$. Analogously, for any buyer $j \in [\ell]$, let $T^{-j}$ denote the joint type space of all buyers other than $j$,
$$T^{-j} := \prod_{j \in [\ell] \setminus \{j\}} [n_j],$$
where an element $\mathbf{-i} \in T^{-j}$ denotes a vector of types for each buyer other than $j$. For any type vector $\mathbf{i} \in T$, suppose that $p_\mathbf{i}$ denotes the probability that $\mathbf{i}$ is realized according to the common prior. Finally, for any buyer $j \in [\ell]$, type $i \in [n_j]$, and good $k \in [m]$, suppose that $v_{\mathbf{i}, j, k}$ is the value that type $i$ of buyer $j$ has for good $k$.

Given this information, a selling mechanism is fully specified the price $r_{\mathbf{i}, j}$ that each buyer $j$ pays when types are revealed according to $\mathbf{i}$, and the quantity $q_{\mathbf{i}, j, k}$ of each good $k$ that each buyer $j$ is allocated when types are revealed according to $\mathbf{i}$. The optimal $\mathbf{r}$ and $\mathbf{q}$ are determined by the following linear program, which maximizes expected revenue subject to individual rationality (IR) and incentive compatibility (IC) constraints.

\lpmax{\displaystyle \sum_{\mathbf{i} \in T} \sum_{j \in [\ell]} p_{\mathbf{i}} r_{\mathbf{i}, j}}{
    \\
    $r_{\mathbf{i}, j} \geq 0$ & for all $\mathbf{i} \in T$, $j \in [\ell]$\\
    $q_{\mathbf{i}, j, k} \geq 0$ & for all $\mathbf{i} \in T$, $j \in [\ell]$, $k \in [m]$\\
    $\displaystyle\sum_{j \in [\ell]} q_{\mathbf{i}, j, k} \leq 1$ & for all $\mathbf{i} \in T$, $k \in [m]$\\
    $\displaystyle\sum_{k \in [m]} v_{\mathbf{i}, j, k} q_{\mathbf{i}, j, k} - r_{\mathbf{i}, j} \geq 0$ & for all $\mathbf{i} \in T$, $j \in [\ell]$ (IR)\\
    $\displaystyle\sum_{k \in [m]} v_{(i, \mathbf{-i}), j, k} q_{(i, \mathbf{-i}), j, k} - r_{(i, \mathbf{-i}), j} \geq \displaystyle\sum_{\mathbf{-i} \in T^{-j}}$ & \raisebox{-.4cm}{for all $j \in [\ell]$, $\mathbf{-i} \in T^{-j}$,}\\
    $p_{(i, \mathbf{-i})}\left(\displaystyle\sum_{k \in [m]} v_{(i', \mathbf{-i}), j, k} q_{(i', \mathbf{-i}), j, k} - r_{(i', \mathbf{-i}), j}\right)$ & \raisebox{.45cm}{$i, i' \in [n_j]$ (IC)}\\
}

\noindent Note that the optimal solution may involve non-integral values of $q_{\mathbf{i}, j, k}$, which can be interpreted as giving good $k$ to buyer $j$ with probability $q_{\mathbf{i}, j, k}$ when types are realized according to $\mathbf{i}$. Also, a more general version of this LP would allow for ex post violations of IR or IC; the more constrained version above is equivalent in terms of the optimal value and resulting buyer surplus, and it produces more reasonable-looking mechanisms.

We implemented an algorithm to compute the optimal mechanism with respect to any seller belief distribution, using Gurobi~\citep{Gurobi} to solve the linear program, with the secondary optimization objective of maximizing buyer surplus. For any prior distribution, we can enumerate over all tuples of partitions of buyer types and compute expected buyer/seller utilities in the corresponding LP for each possible vector of disclosure messages, then combine to compute the overall expected utilities of the disclosure game. We ran thousands of trials with valuations and probabilities drawn independently and uniformly at random from $[0, 1]$; the counterexamples in Sections \ref{subGeneralImpossibilityMultipleGoods} and \ref{subGeneralImpossibilityMultipleBuyers} were discovered in this process, then subsequently simplified.

\subsection{Multiple goods}\label{subGeneralImpossibilityMultipleGoods}

Surprisingly, even for the simple case of \textbf{one buyer}, \textbf{two goods}, and two buyer types, it is possible for the partition $\{\{1, 2\}\}$ (in which there is no voluntary disclosure) to be the best partition (and thus the best pure-strategy equilibrium) in terms of ex ante buyer utility, even though it results in the buyer sometimes not getting one of the goods. The following is a simple example in which the valuations of the two goods are correlated.

\begin{center}
	\begin{tabular}{r|c|c|c}
		Type & Probability & Value for good 1 & Value for good 2 \\\hline
		1 & $1/2$ & 3 & 4 \\\hline
		2 & $1/2$ & 4 & 9
	\end{tabular}
\end{center}
In the no-disclosure equilibrium induced by the partition $\{\{1, 2\}\}$, the unique optimal mechanism is for the seller to post the following menu of choices.
\begin{center}
	\begin{tabular}{c|c}
		Bundle & Price \\\hline
		Only good 1 & 3 \\\hline
		Both goods 1 and 2 & 12
	\end{tabular}
\end{center}

Type 1 buyers purchase only good 1, and type 2 buyers purchase both goods 1 and 2. Notice that type 2 buyers get utility $1 > 0$. This is because the seller is unable to extract any more utility from them, for otherwise, they would opt to only buy good 1 (and if the seller raised the price on good 1, they would completely exclude type 1 buyers, hurting revenue even more). There is some inefficiency though, as good 2 is only sold with ex ante probability $\frac12$, even though the buyer always has positive utility for it. The only other partition to consider is $\{\{1\}, \{2\}\}$, in which the buyer exactly reveals their type. While this always yields an efficient outcome, the seller is clearly able to extract all of the surplus, leaving the buyer with utility zero. Thus, we conclude that the Efficiency Lemma no longer holds when there are two goods.

In fact, it turns out that this can happen even when the valuations for the two goods are independent, as shown in the following example (which has been simplified as much as possible).

\begin{center}
	\begin{tabular}{r|c|c|c}
		Type & Probability & Value for good 1 & Value for good 2 \\\hline
		1 & $0.15 \cdot 0.4$ & 56 & 38 \\\hline
		2 & $0.15 \cdot 0.6$ & 56 & 69 \\\hline
		3 & $0.85 \cdot 0.4$ & 91 & 38 \\\hline
	    4 & $0.85 \cdot 0.6$ & 91 & 69
	\end{tabular}
\end{center}
In this case, in the unique optimal mechanism is the following.
\begin{center}
	\begin{tabular}{c|c}
		Bundle & Price \\\hline
		Good 2 for sure, and good 1 with probability $31/35$ & 118.6 \\\hline
		Both goods 1 and 2 & 129
	\end{tabular}
\end{center}

A type 1 buyer will purchase nothing, a type 2 buyer will purchase the randomized bundle, and types 3 and 4 purchase both goods. Note that the $31/35$ is the maximum probability at which types 3 and 4 weakly prefer buying both goods. Only type 4 gets nonzero utility. By exhaustively checking each of the 14 alternative partitions of $\{1, 2, 3, 4\}$, we verified that, in any pure-strategy equilibrium in which both goods always get sold, the ex ante buyer utility is strictly lower.

\subsection{Multiple buyers}\label{subGeneralImpossibilityMultipleBuyers}

Now suppose there is \textbf{one good} and \textbf{two buyers}, whose valuations for the good are independent and identically distributed as follows.

\begin{center}
	\begin{tabular}{r|c|c}
		Type & Probability & Value for good \\\hline
		1 & $1/4$ & 1 \\\hline
		2 & $1/4$ & 2 \\\hline
		3 & $1/2$ & 3
	\end{tabular}
\end{center}

In the absence of disclosure, an optimal mechanism for the seller (as verified computationally) is as follows. If at least one buyer bids 3, randomly choose one such buyer and sell them the good for 2.5. If no buyer bids 3, but at least one buyer bids 2, randomly choose one such buyer and sell them the good for 2. Otherwise, do not sell the good. The expected buyer surplus is $\frac34 \cdot \frac12 = \frac38$, as there is a $\frac34$ chance that there will be some buyer of type 3, in which case the buyer surplus will be $\frac12$. However, the no-disclosure equilibrium is clearly not efficient, as the good is not sold $\frac{1}{16}$ of the time (when both buyers have type 1).

An intuitive idea to repair this equilibrium so that it is efficient is to use the partition $\{\{1\}, \{2, 3\}\}$. By separating the lowest-value type from the other types, we can prevent the inefficient outcome since there is no longer any incentive for the seller to refuse to sell. In fact, in the one-buyer case, this partition is precisely the one constructed in the proof of the Efficiency Lemma. In this case, when the buyer sends the message $\{1\}$, the seller sells the good for a price of 1, and efficiency is regained with no loss to buyer surplus; and when the buyer sends the message $\{2, 3\}$, the seller's incentives are the same as in the original equilibrium, so again, there is no loss to buyer surplus.

However, with two buyers, the seller's incentives actually \emph{do} change, and for this particular example, they change in a way that \emph{harms} buyer surplus. In the partitional equilibrium where both buyers use the partition $\{\{1\}, \{2, 3\}\}$ and both send the message $\{2, 3\}$, the optimal mechanism is the same as in the no-disclosure equilibrium, except that the price the seller charges to a buyer who bid 3 is 2.75 instead of 2.5, and hence the buyer surplus is $\frac89 \cdot \frac14 = \frac29 < \frac38$. Even worse, when one buyer sends the message $\{1\}$ and the other buyer sends the message $\{2, 3\}$, having the outside option of selling the good to the first buyer for 1 strictly incentivizes the seller to only sell to the second buyer for a price of 3. This means that the buyer surplus is zero, and the good might be sold to a buyer of value 1 over a buyer of value 2. So overall, using the partition $\{\{1\}, \{2, 3\}\}$ for both buyers results in a lower surplus of $\frac34 \cdot \frac34 \cdot \frac29 = \frac18$, yet still does not resolve the inefficiency. In fact, the inefficiency is even greater, as the seller's welfare gains do not offset the buyers' welfare losses.

Having only one buyer use the partition $\{\{1\}, \{2, 3\}\}$ and the other buyer use the no-disclosure partition $\{\{1, 2, 3\}\}$ suffers from similar problems. While the total social welfare is the same, the buyer surplus is lower, and the equilibrium is still inefficient. By enumerating over all pairs of partitions of the set $\{1, 2, 3\}$, we verified that, despite its inefficiency, the no-disclosure equilibrium yields the strictly highest expected buyer surplus. This concludes the proof of Theorem \ref{thmGeneralImpossibility}.


\section{Complexity of welfare maximization}\label{secComplexity}

In settings where disclosure schemes \emph{can} improve buyer welfare, there arises the computational question of how to find such schemes. In this section we study the complexity of finding a buyer-optimal pure-strategy equilibrium over a discrete distribution. We assume that there is one good and one buyer with possible types $N = \{1, \dots, n\}$, each with value $v_1 < \ldots < v_n$ and occurring with probability $p_1, \ldots, p_n$ respectively. 

First, we analyze the restricted setting introduced by~\citep{ali20voluntary}, in which buyer messages must be connected, and thus the search problem is restricted to connected partitions. Note that Example 1 of~\citep{ali20voluntary} illustrates that their greedy algorithm is unable to compute the buyer-optimal disclosure strategy in this setting. Using dynamic programming, we give a polynomial time algorithm to solve this problem.

Next, we ask whether the buyer could do even better if they were allowed to report from a broader class of messages. Specifically, we turn our attention to a more general setting in which a buyer may report any arbitrary subset of their values $\{v_1, \ldots, v_n\}$ (or, equivalently, an arbitrary subset of $N$). Here, things are not quite as easy: it is weakly NP-hard to compute the buyer-optimal disclosure scheme. Further, the optimal surplus using connected messages does not even provide a constant-factor approximation to the unconstrained optimal buyer surplus.

\subsection{Connected partitions}\label{subComplexityConnected}

Let us first consider the case of connected messages. By the Partitional Lemma, it suffices to restrict attention to equilibria induced by some partition $\mathcal{P}$ of the set of buyer types $N = \{1, \dots, n\}$. We show the following.

\begin{theorem}\label{thmDP}
    \Cref{alg:opt-connected} computes a connected partition that induces an equilibrium which maximizes ex ante buyer utility in polynomial time.
\end{theorem}

\begin{algorithm}[H]
	\SetAlgoLined
	\KwIn{Values $v_1 < \cdots < v_n$ and probabilities $p_1, \ldots, p_n$}
	\KwOut{A connected partition inducing an equilibrium with optimal ex ante buyer utility}
	Initialize array $Q[0\ldots n]$\;
	$Q[0].\texttt{utility} \gets 0$\;
	$Q[0].\texttt{partition} \gets \emptyset$\;
	\For{$i \gets 1 \ldots n$}{
	    $j^* = \argmax_{j \in \{0, \ldots, i - 1\}} \{Q[j].\texttt{utility} + \texttt{buyerUtility}(\{j+1, \ldots, i\})\}$\;
	    $Q[i].\texttt{partition} = Q[j^*].\texttt{partition} \cup \{\{j + 1, \ldots, i\}\}$\;
	    $Q[i].\texttt{utility} = Q[j^*].\texttt{utility} + \texttt{buyerUtility}(\{j + 1, \ldots, i\})$\;
	}
	\Return $Q[n].\texttt{partition}$\;
	\caption{Optimizes buyer welfare over connected partitions.}
	\label{alg:opt-connected}
\end{algorithm}

\begin{proof}
    First we will prove the correctness of \Cref{alg:opt-connected}. It uses a dynamic programming approach with array $Q$. Each $Q[i]$ will store the best partition of $\{1, \ldots, i\}$ as well as the expected buyer utility when using this partition, scaled down by $\sum_{j=1}^i p_i$. Note that when we are considering the next element $i$, conditioned on it being in some partition element $\{j + 1, \ldots, i\}$, the remaining elements of the partition will be the optimal partition of $\{1, \ldots, j\}$. Hence, by choosing the best possible $j$, that will allow us to find the optimal partition for $Q[i]$. The \texttt{buyerUitility} function, given a possible message of the buyer, gives the highest utility the buyer can receive when the seller best responds to this message.
        
    To show the runtime is polynomial, note that the initialization can be done in $O(n)$ time. The for loop is runs for $n$ iterations. In each iteration, note that \texttt{buyerUitility} can be computed by checking each seller price in $\{v_j, \ldots, v_i\}$, of which there are at most $n$. For each of these, we must check the probability of sale and expected buyer utility. This can be done in time linear in $n$ by iterating through the values in order, and keeping track of the total probability the buyer has value above this value. In total, this loop takes $O(n^3)$ time.
\end{proof}

\subsection{Arbitrary partitions}\label{subComplexityArbitrary}

While message connectivity might reflect realistic practical constraints on the set of feasible messages, it is nevertheless interesting to consider settings where this constraint is not present, and so the buyer could potentially do even better. Unfortunately, it turns out that, in the general case, computing a pure-strategy equilibrium that maximizes ex ante buyer utility is NP-hard. More formally, we consider the following decision problem.

\textbf{BUYER-OPT:} \emph{Given a sequence of probabilities $\seq{p}{n}$ (where $\sum_{i \in [n]}p_i = 1$), corresponding distinct positive valuations $\seq{v}{n}$, and a positive number $U$, determine whether there exists a pure-strategy equilibrium with expected buyer utility $U$ in the disclosure game with one buyer, one seller, and one good, where the buyer has value $v_i$ for the good with probability $p_i$.}

\begin{theorem}\label{thmHardness}
    BUYER-OPT is weakly NP-complete.
\end{theorem}

\ipnc{.55}{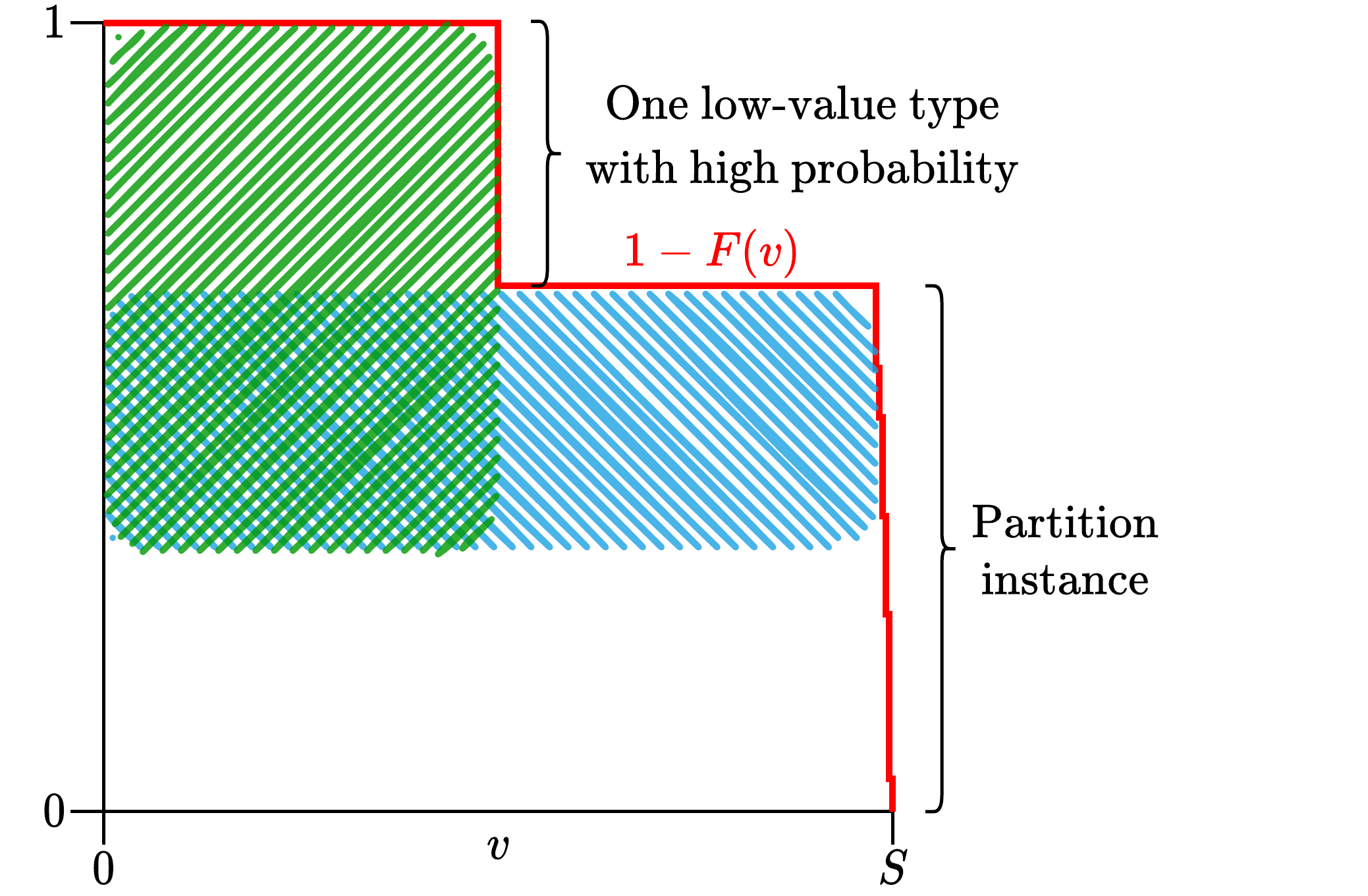}{\label{figReduction}The reduction from PARTITION to BUYER-OPT.}

As might be expected, we reduce from the PARTITION problem, defined as follows.

\textbf{PARTITION:} \emph{Given a finite sequence of positive integers $\seq{s}{m}$, adding up to some even positive integer $S$, determine whether there exists $I \subseteq [m]$ such that
$$\sum_{i \in I}s_i = \frac{S}{2}.$$}

The main idea behind the reduction is that there will be one low-value type with high probability and many high value types with low probability, as shown in \Cref{figReduction}. The zigzagging red line shows the probability of sale given a posted price according to the instance defined by the reduction. Ideally, to maximize buyer welfare, we would like to pool some of the high-value types together with the low-value type, so that all types in the pool send the same message and the seller will set their price equal to the value of the low-value type, giving the high-value types positive utility. The seller maximizes revenue by choosing the largest rectangle under the curve, ignoring the types that are not in this pool. Thus, in order to incentivize the seller to choose the low price, we need the vertical green rectangle in \Cref{figReduction} to be larger than the horizontal blue rectangle, which is only possible if we pool at most half of the high-value types (weighted by probability mass) with the low-value type. Optimally, we would like to get exactly half of the probability mass, which requires solving PARTITION. See \Cref{appHardnessProof} for the full proof.

Finally, we now ask whether our polynomial time algorithm from Section \ref{subComplexityConnected} could be used to at least approximate optimal ex ante buyer utility in this more general setting. We answer this question in the negative.

\begin{proposition}\label{proNoApprox}
    No equilibrium for the general setting induced by a connected partition can have ex ante buyer utility that approximates the optimal within a constant factor.
\end{proposition}

See \Cref{appProNoApproxProof} for the proof.


\section{Conclusion}\label{secConclusion}

Can buyers reveal for a better deal, and if so, how? In the restricted setting with one buyer and one good, \citet*{ali20voluntary} answer the former question with an emphatic ``Yes!'' Utility-improving disclosure schemes are generally feasible, always socially efficient, and in the presence of connectivity constraints on the disclosure technology, can easily be exactly optimized for the buyer. Unfortunately, it appears that these desirable properties are not very robust.

We believe that our negative results are not mere peculiarities of the model but are instead due to fundamental market forces. With either multiple goods or multiple buyers, in the absence of disclosure, the buyer(s) benefit substantially from the uncertainty the seller faces. While, at the surface, it still seems plausible that a buyer should only gain utility from credibly revealing to the seller that they would not be willing to pay the seller's optimal reserve price, such disclosures may have secondary effects, dramatically reducing the seller's uncertainty to the point where buyers are ultimately harmed in expectation.

We do note that, despite all of this negative evidence on the benefit of disclosure, there are, in fact, instances where it can provably help, even in the more complicated settings we consider.\footnote{As an example, consider the following alternative prior distribution in which buyers have i.i.d.\@ values, symmetric strategies, and both Pareto improve in terms of interim utility by disclosing some information. The value distribution will be discrete; with probability $\frac{99}{100}$ they will have value $1000$ and with probability $\frac{1}{200}$ each, they will have value $1$ and $2$. With no disclosure, the revenue-optimal mechanism run by the seller will only sell at a price of $1000$. This means that buyers never receive positive utility as they cannot purchase the good for a price strictly less than their value. On the other hand, suppose buyers disclose whether they are of ``low'' type with value in $\{1, 2\}$ or ``high'' type with value $1000$. In the scenario where both buyers reveal a low type, from the seller's perspective, each buyer has value $1$ or $2$ with probability $\frac{1}{2}$, so the optimal selling mechanism is a second price auction with reserve price $1$. Hence, in the case where one buyer has value $2$ and the other has value $1$, the former will receive positive utility, strictly improving on the no-disclosure setting..}
We argue only that the compatibility of buyer and social welfare and the buyer advantages of disclosure are no longer guaranteed in larger, realistic market settings.
Depending on the setting, instances with multiple buyers/goods where disclosure is useful may well be the exception rather than the norm.

There are still many dimensions to the voluntary disclosure model that have yet to be explored. Most notable is the possibility of mixed strategies. A tantalizing open question, left unaddressed by this as well as prior work, is whether there exist mixed-strategy equilibria in which the buyer(s) obtain strictly higher ex ante utility than in any pure-strategy equilibrium---one could imagine a scenario where some buyer types are indifferent between multiple messages, some seller types are indifferent between multiple prices, and somehow the complex belief distributions generated yield a higher expected buyer surplus than would be possible with those generated by pure messaging strategies. Even with the restriction that there is one buyer and one good, we do not see an obvious way to rule out this possibility (even for simple distributions like $U[0, 1]$), and while such a scenario may sound absurd in the one-buyer, one-good setting, it seems entirely plausible with multiple buyers or goods.

An orthogonal extension of the model would be to allow the buyers to \emph{privately} disclose information to the seller, without other buyers observing the value. While possibly more applicable in some scenarios, this model appears to be less tractable, as the seller's revenue maximization problem cannot be solved independently for every possible tuple of buyer messages. Since the buyers face uncertainty over which subgame they are in, standard tools from auction theory do not apply; e.g., even for the setting from \Cref{secMultipleUniform01Buyers} of two uniform $[0, 1]$ buyers sending connected messages, a Myerson auction is no longer guaranteed to be optimal or incentive compatible.

A final direction for future work concerns the complexity of the BUYER-OPT decision problem. Our reduction is from PARTITION, which is weakly NP-Hard, and it is clear that the optimal values of the instances produced by the reduction can be well-approximated in polynomial time. This leaves open the possibility of a pseudo-polynomial time algorithm or a polynomial time approximation scheme for solving this problem.

\section*{Acknowledgments}

We are deeply grateful to Brendan Lucier and Nicole Immorlica for their generous advice and feedback on earlier versions of this paper. We would also like to thank our anonymous IJCAI reviewers for many useful suggestions on improving clarity.

This material is based upon work supported by the National Science Foundation Graduate
Research Fellowship Program under Grant No. DGE1745303. Any opinions, findings, and conclusions or recommendations expressed in this material are those of the author(s) and do not necessarily reflect the views of the National Science Foundation.

\bibliographystyle{plainnat}
\bibliography{abb,bibliography}

\onecolumn
\newpage
\appendix
\section*{Appendix}

\section{Analysis of Single-Threshold Symmetric Disclosure} \label{appOneSixth}

We show that for independent $v_A, v_B \sim U[0,1]$, if both buyers disclose according to the high/low partition $\{[0,t], [t,1]\}$ for $t \in [0, 1/2]$, then the expected buyer surplus is exactly $1/6$.
This claim and this calculation are a straightforward generalization of \Cref{subMultipleBuyersDiabolo}.

\medskip 
Since the buyers' distributions and disclosure schemes are symmetric, it suffices to calculate the surplus for buyer $A$. 
There are four cases, corresponding to the possible pairs of disclosures $\{[0,t], [t,1]\}^2$.
If $(v_A, v_B) \in [0,t]^2$, then it is straightforward to show that the seller sets a reserve price of $t/2$ and runs a second-price auction. 
Splitting the region into the portion where $A$ pays the reserve price and the region where $A$ pays $v_B$, the surplus for buyer $A$ is then
\[
    U_A^{L,L} = \int_{t/2}^{t} \int_0^{t/2} v_A - t/2 \:dv_B \: dv_A + \int_{t/2}^{t} \int_{t/2}^{v_A} v_A - v_B \:dv_B \: dv_A = \frac{t^3}{12},
\]
where this integral incorporates the likelihood that $(v_A,v_B)$ are in this region.
Similarly, if $(v_A, v_B) \in [t,1]^2$, then the seller runs a second-price auction with a reserve price of $1/2$. 
The surplus for buyer $A$ is then
\[
    U_A^{H,H} = \int_{1/2}^{1} \int_{1/2}^{v_A} v_A - v_B \:dv_B \: dv_A = \frac{1}{12} - \frac{t}{8}.
\]

It remains to calculate the surplus for $A$ in the case when $(v_A, v_B) \in [0,t]\times [t,1]$ and when $(v_A, v_B) \in [t,1]\times [0,t]$.
Rather than analyze the Myerson auction directly, we may leverage the calculations developed in Section \ref{subMultipleGoodsBuyersPartitions}, in particular \cref{eq:buyersurplussimplification}, in order to find the expected surplus for buyer $A$ in each of these regions:
\[
    U_A^{H,L} = \int_{1/2}^{1/2 + t/2} \int_{0}^{v_A - (1-t)/2} 1 - v_A \: dv_B \: dv_A + \int_{1/2 + t/2}^1 \int_{0}^t 1 - v_A \: dv_B \: dv_A = \frac{t}{8} - \frac{t^2}{16} + \frac{t^3}{48},
\]
and 
\[
    U_A^{L,H} = \int_{t/2}^t \int_t^{x + (1-t)/2} t - x \: dv_B \: dv_A = \frac{t^2}{16} - \frac{5t^3}{48}.
\]
Summing these together, we find that total expected surplus for buyer $A$ is $U_A =U_A^{L,L} + U_A^{H,H} + U_A^{H,L} + U_A^{L,H} = 1/12$, and so by symmetry the total expected buyer surplus is $1/6$, as claimed.

\section{Proof of \Cref{thmHardness}}\label{appHardnessProof}

    To see that BUYER-OPT is in NP, observe that the Partitional Lemma implies there is a buyer-optimal equilibrium induced by a partition of $N$. Given such a partition $\mathcal{P}$, for every $P \in \mathcal{P}$, there are at most $\abs{P}$ prices the seller may choose to set (a price not equal to $v_i$ for some $i \in P$ cannot be optimal). Therefore, we can efficiently compute the optimal selling mechanism and its expected buyer utility, then verify that the utility is at least $U$.
    
    To prove hardness, we reduce from PARTITION, which is known to be weakly NP-hard~\citep{KarpNPCompleteProblems}. In terms of the notation defining BUYER-OPT and PARTITION, our reduction is defined as follows:
    \newpage
	\begin{align*}
		n &:= m + 1\\
		p_i &:= \frac{2s_i}{3S} & \txt{for } i \leq m\\
		v_i &:= S - \frac{1}{4 + i} & \txt{for } i \leq m\\
		p_n &:= \frac13\\
		v_n &:= \frac{S}{2}\\
		U &:= \frac{S}{6} - \frac{1}{12}
	\end{align*}

	For the forward direction, suppose that $I$ is a solution to the given PARTITION instance. Then consider the partitional equilibrium induced by the partition
	$$\{I \cup \{n\}, [m] \setminus I\}.$$
	To compute the buyer surplus, there are three cases to consider:
	\begin{enumerate}
		\item\label{itmBuyerTypeN} The buyer's type is $n$.
		\item\label{itmBuyerTypeI} The buyer's type is in $I$.
		\item\label{itmBuyerTypeMMinusI} The buyer's type is in $[m] \setminus I$.
	\end{enumerate}
	Observe that the first case occurs with probability $\frac13$, and the second case occurs with probability
	$$\sum_{i \in I} p_i = \sum_{i \in I} \frac{2s_i}{3S} = \frac{2\sum_{i \in I} s_i}{3S} = \frac{2 \cdot \frac{S}{2}}{3S} = \frac13,$$
	so, in fact, all three cases occur with probability $\frac13$. Consider the utility of the buyer in case (\ref{itmBuyerTypeI}), in which the buyer reveals that their type is in $I \cup \{n\}$, so the seller knows they are in case (\ref{itmBuyerTypeN}) or (\ref{itmBuyerTypeI}). Given this information, with probability $\frac12$ we are in case (\ref{itmBuyerTypeN}), so the buyer's value is $\frac{S}{2}$, and with probability $\frac12$, we are in case (\ref{itmBuyerTypeI}), so the buyer's value is in the range $(S - \frac{1}{4}, S)$. If the seller sets a price in the range $(\frac{S}{2}, S)$, they will exclude type $n$ and only make a sale with probability $\frac12$, so their expected revenue is strictly less than $\frac{S}{2}$. If the seller sets a price of $S$ or greater, they will definitely not make a sale, so their expected revenue is zero. In either case, their expected revenue is less than $\frac{S}{2}$, which is what they could get by setting a price of $\frac{S}{2}$ and guaranteeing sale. Therefore, the seller's optimal price is at most $\frac{S}{2}$. Since we are in case (\ref{itmBuyerTypeI}), the buyer has value at least $S - \frac14$, so the expected buyer surplus will be at least
	$$S - \frac14 - \frac{S}{2} = \frac{S}{2} - \frac14.$$
	Since case (\ref{itmBuyerTypeI}) occurs with probability $\frac13$, the overall expected buyer surplus is at least
	$$\frac13\left(\frac{S}{2} - \frac14\right) = U.$$
	
	For the backward direction, suppose there is a pure-strategy equilibrium giving the buyer ex ante expected utility at least $U$. By the Partitional Lemma, it is without loss of generality to assume the equilibrium is induced by some partition $\mathcal{P}$ of $[n]$. Let $I \subseteq [m]$ be the unique set of types such that $I \cup \{n\} \in \mathcal{P}$. We claim that $I$ is a solution to the PARTITION instance, i.e.,
	$$\sum_{i \in I} s_i = \frac{S}{2}.$$
	
	For any $i \in [n]$, let $r_i$ be the price the seller charges type $i$ at equilibrium. Then
	\begin{align*}
		\frac{S}{6} - \frac{1}{12} &= U\\
		&\leq \sum_{i \in [n]} p_i(v_i - r_i) \stext{by the Efficiency Lemma and the definition of $U$}\\
		&= \sum_{P \in \mathcal{P} \setminus \{I \cup \{n\}\}} \sum_{i \in P} p_i(v_i - r_i) + \sum_{i \in I} p_i(v_i - r_i) + p_n(v_n - r_n)\\
		&= \sum_{P \in \mathcal{P} \setminus \{I \cup \{n\}\}} \sum_{i \in P} p_i \left(v_i - \min_{j \in P} v_j\right) + \sum_{i \in I} p_i(v_i - v_n) + p_n (v_n - v_n) \stextn{by the Efficiency Lemma}\\
		&< \sum_{P \in \mathcal{P} \setminus \{I \cup \{n\}\}} \sum_{i \in P} p_i \left(S - \left(S - \frac14\right)\right) + \sum_{i \in I} p_i(S - v_n) \stextn{since, for all $i \in [m]$, $v_i \in (S - \frac{1}{4}, S)$}\\
		&= \sum_{P \in \mathcal{P} \setminus \{I \cup \{n\}\}} \sum_{i \in P} p_i \left(\frac{1}{4}\right) + \sum_{i \in I} p_i\left(S - \frac{S}{2}\right)\\
		&= \frac{1}{4} \sum_{P \in \mathcal{P} \setminus \{I \cup \{n\}\}} \sum_{i \in P} p_i + \sum_{i \in I} \frac{2s_i}{3S} \cdot \frac{S}{2}\\
		&= \frac{1}{4} \sum_{P \in \mathcal{P} \setminus \{I \cup \{n\}\}} \sum_{i \in P} p_i + \frac13 \sum_{i \in I} s_i\\
		&\leq \frac14 \cdot \frac23 + \frac13 \sum_{i \in I} s_i\\
		&= \frac16 + \frac13 \sum_{i \in I} s_i
	\end{align*}
	Multiplying by 3 and rearranging, we have
	$$\sum_{i \in I} s_i > \frac{S}{2} - \frac34.$$
	Since $\sum_{i \in I} s_i$ and $\frac{S}{2}$ are integers, it follows that
	$$\sum_{i \in I} s_i \geq \frac{S}{2}.$$
	
	To prove the opposite inequality, suppose that the buyer reveals that their type is in $I \cup \{n\}$. If the seller sets a price of $S - \frac{1}{4}$, they exclude type $n$, and thus sell with probability
	\begin{align*}
		\frac{\sum_{i \in I} p_i}{\frac13 + \sum_{i \in I} p_i},
	\end{align*}
	so their expected revenue is
	$$\left(S - \frac{1}{4}\right)\left(\frac{\sum_{i \in I} p_i}{\frac13 + \sum_{i \in I} p_i}\right).$$
	Since the Efficiency Lemma implies that the seller instead chooses to sell at a price of $\frac{S}{2}$ for an expected revenue of $\frac{S}{2}$, we must have that
	$$\left(S - \frac{1}{4}\right)\left(\frac{\sum_{i \in I} p_i}{\frac13 + \sum_{i \in I} p_i}\right) \leq \frac{S}{2},$$
	or, equivalently,
	$$\left(S - \frac{1}{4}\right) \sum_{i \in I} p_i \leq \frac{S}{2} \left(\frac13 + \sum_{i \in I} p_i\right).$$
	Rearranging again, we have
	$$\sum_{i \in I} p_i \leq \frac{\frac{S}{2} \cdot \frac13}{S - \frac14 - \frac{S}{2}} = \frac{\frac{S}{6}}{\frac{S}{2} - \frac14} = \frac{1}{6S - 3} + \frac13.$$
	Plugging in the definition of $p_i$, this becomes
	$$\frac{2}{3S} \sum_{i \in I} s_i \leq \frac{1}{6S - 3} + \frac13,$$
	which simplifies to
	$$\sum_{i \in I} s_i \leq \frac{3S}{2(6S - 3)} + \frac{3S}{6} = \frac{S}{4S - 1} + \frac{S}{2}$$
	Since $\sum_{i \in I} s_i$ and $\frac{S}{2}$ are integers and $\frac{S}{4S - 1} < 1$ for any positive integer $S$, it follows that
	$$\sum_{i \in I} s_i \leq \frac{S}{2}.$$
	Thus, we have equality, so $I$ is a solution to the PARTITION instance. \qed

\section{Proof of \Cref{proNoApprox}}\label{appProNoApproxProof}
    
    Consider a setting in which a buyer has possible values $1, 2, 2 + \delta$ for some $0 < \delta < 1$ each with probabilities $\frac{1}{3}, \frac{5}{9}, \frac{1}{9}$ respectively. We will show that every equilibrium induced by a connected partition gives the buyer expected utility at most $\frac{\delta}{9}$ while an arbitrary partition can give expected utility at least $\frac{1 + \delta}{9}$. This implies no approximation ratio better than $\frac{1 + \delta}{\delta}$ is achievable. By letting $\delta$ approach $0$, this lower bound grows arbitrarily large.
    
    Let us first show that when a buyer is allowed to report arbitrary subsets, there are equilibria in which the buyer ex ante utility is at least $\frac{1 + \delta}{9}$. Consider the equilibrium induced by the partition $\{\{1, 2 + \delta\}, \{2\}\}$. Now, when the buyer reports $\{1, 2 + \delta\}$, the seller will prefer to set a price of $1$, which will result in expected revenue $1$, over $2 + \delta$, which (after updating their posterior,) will only sell with probability $\frac{1}{4}$, resulting in expected revenue $\frac{2 + \delta}{4} < 1$. Hence, by setting a price of $1$, the buyer receives utility $1 + \delta$ when they have value $2 + \delta$, so their ex ante utility is at least $\frac{1 + \delta}{9}$.
    
    Next, we will upper bound the ex ante buyer utility of a connected partition strategy. First, note that a seller will always choose a posted price $p \in \{1, 2, 2 + \delta \}$. This implies that whenever a buyer has value $1$, they cannot receive any utility, as the price will always be at least $1$. Further, we will show that when a buyer has value $2$, the seller will always set a posted price of at least $2$. Indeed, regardless of what the buyer reports, since messages must contain the buyer's true value, the seller knows that $2$ is a possibility. Hence, if the buyer posts a price of $1$, they will get expected revenue $1$, while if they post a price of $2$, they will get expected revenue at least $\frac{5}{9} \cdot 2 > 1$. This implies in equilibrium they will never set a price of $1$, so they buyer cannot receive any utility. Finally, let us consider what happens when the buyer has true utility $2 + \delta$. In this case, if they report $\{2 + \delta \}$, the seller will set the price to $2 + \delta$ and they will receive zero utility. On the other hand, any other report will include $2$ in the message, and by previous argument, the seller will not set a price of $1$, so the most utility they can receive is $2 + \delta - 2 = \delta$. Hence, the buyer's ex ante utility is at most $\frac{\delta}{9}$. \qed
\end{document}